\documentclass[preprint2]{aastex}
\def\ha{H${\alpha}$}
\def\hb{H${\beta}$}

\received{}
\shortauthors {Shaw et al.}
\shorttitle {Morphology and Evolution of LMC Planetary Nebulae} 

\begin{document}

\title {Morphology and Evolution of the LMC Planetary Nebulae\footnote
{Based on observations with the NASA/ESA Hubble Space Telescope, 
obtained at the Space Telescope Science Institute, which is operated 
by the Association of Universities for Research in Astronomy, Inc., under
NASA contract NAS 5-26555.} 
}

\author {Richard A.~Shaw, Letizia Stanghellini\altaffilmark{2}, 
Max Mutchler}
\affil {Space Telescope Science Institute, Baltimore, MD 21218}
\email {shaw@stsci.edu, lstanghe@stsci.edu, mutchler@stsci.edu}
\author {Bruce Balick}
\affil {University of Washington, Seattle, WA 98195}
\email {balick@astro.washington.edu}
\altaffiltext{2}{Affiliated with the Astrophysics Division, Space Science
Department of ESA; on leave from Osservatorio Astronomico di Bologna}

\and
\author {J.~Chris Blades}
\affil {Space Telescope Science Institute, Baltimore, MD 21218}
\email {blades@stsci.edu}

\clearpage

\begin{abstract} 

The LMC is ideal for studying the co-evolution of planetary nebulae (PNe) 
and their central stars, in that the debilitating uncertainties of the 
Galactic PN distance scale and selection biases from attenuation 
by interstellar dust do not apply.  We present images and analyze 
slit-less spectra which were obtained in a survey of Large Magellanic 
Cloud PNe.  These data on 29 
targets were obtained with {\it HST} using the Space Telescope Imaging 
Spectrograph.  The data permit us to determine the nebular dimensions 
and morphology in the monochromatic light of several emission 
lines, including those that have traditionally been used for 
morphological studies in the Galaxy: H$\alpha$, [\ion{N}{2}] 6583~\AA\ 
and [\ion{O}{3}] 5007~\AA, plus others of varying ionization, 
including [\ion{O}{1}], \ion{He}{1}, and [\ion{S}{2}].  
Together with the 31 resolved LMC PNe for which monochromatic images exist 
in the {\it HST} archive, these data show that the incidence of 
non-symmetric nebulae, including bipolar nebulae (which is an indicator 
of Population I ancestry in the Galaxy), is significantly higher than that 
reported for the Galaxy.  
The onset of asymmetric features appears even in very young nebulae (with 
dynamical ages of $\sim1400$ yr), suggesting that at least the gross 
features of the nebular morphology may be more closely tied to PN formation, 
and that 
subsequent shaping of the expanding envelope by the radiation field and 
wind from the central star may play the lesser role of amplifying these 
gross features.  There is some evidence of evolution between two 
morphological types, in the sense that bipolar core (BC) nebulae may 
evolve to pure bipolars late in the PN lifetime. 

\end{abstract}


\section {Introduction} 

The nature of the physical connection between planetary nebulae (PNe) and 
their central stars has been studied for the better part of the last century, 
and a good deal of physical insight into the late stages of stellar evolution 
has emerged as a consequence.  As images have improved, it has become 
surprisingly clear that mass loss is far from isotropic.  These and other 
observations have challenged our understanding 
of the formation of PNe, and of the role of their central stars (CSs) in the 
subsequent evolution of the shells.  This latter topic has received a great 
deal of attention in the past three decades, driven by the availability of 
new techniques in observing and in theoretical modelling.  That the shapes 
of the ejected PN shells should somehow relate to the formation mechanism 
is perhaps easy to intuit, but the connection between nebular morphology 
and the Population type of the progenitor stars \citep{Greig72} is not 
so easy to understand.  Subsequent work by many investigators 
\citep[e.g.,][to name but a few]{Peimbert78,PTP83,Kaler83}
began to connect nebular abundances to the results of stellar evolution 
processes in the central stars immediately prior to PN formation.  It is 
primarily from these connections that the relationship between asymmetric 
PNe and higher CS (and progenitor star) mass can be inferred.  

But what is it, exactly, that PN morphologies tell us about their formation 
history?  How can we relate morphology to the physical properties of the 
nebulae, their environments, and especially their CSs?  In principle, 
morphology should reflect one or more factors, which we might characterize 
as either ``local'' --- relating to properties of the CS and the 
immediate vicinity at the time of formation --- or ``non-local'' --- relating 
to conditions that exist or develop after the PN is formed.  Local 
properties would include the dynamics of the gas at the moment of ejection, 
departures from spherical symmetry of the gravitational or magnetic field 
immediately surrounding the CS, and by the amount of dust within the 
expelled gas.  
Non-local properties include the circumstellar environment, including mass 
ejected from the CS prior to the PN ejection and the density of the local 
ISM, and the radiation field and wind from the CS, both of which change 
rapidly as the star evolves to a white dwarf.  
\citet{Balick87} presented compelling evidence for a restricted sample of 
Galactic PNe that nebular morphologies can be greatly influenced by the 
energy (from wind and radiation) deposited in the nebula by the CS. 
What was lacking most was a sense of the relative importance of local 
vs. non-local factors on nebular morphology, and whether the influence of 
these factors varied with morphological type.   
While comprehensive catalogs of PNe have been available for some time 
\citep[e.g.][]{Acker_etal82}, no systematic morphological classification 
scheme on a large PN sample was published before the 1990s. 

In the early part of the last decade \cite{SCM92} published a catalog of 
nearly 300 southern Galactic PNe, and a few years later a similar effort was 
completed by \citet{Manchado_etal96} for 243 northern PNe.   
More recently, \cite{Gorny_etal99} published a catalog of an additional 100 
southern PNe. 
Combined with earlier, more limited image catalogs, these large databases 
are proving essential for defining and refining the morphological classes, 
and for deriving their statistical properties.  
A number of papers based on these catalogs lead to a new view of PN 
formation, and how it is related to the evolution of the progenitor stars 
\citep{SCS93, CS95}. 
The interpretation that emerged was that PN morphology is an essential 
physical property, linked to details of the progenitor formation (i.e., 
the parent stellar population), the main sequence mass and evolution of the 
progenitor, and the dynamical evolution of the nebula once it was formed.  
In particular, it confirmed that progenitors of the most asymmetric PNe 
belong to a younger stellar population than those that form elliptical or 
round PNe, that they are on average more massive and produce more massive
central stars, and that the nebulae are rich in nitrogen but are relatively 
carbon-poor compared with most PNe. 

Many of the important physical properties that one can derive from 
observations of a PN and its central star, including the nebular dimensions, 
mass, dynamical age, and the stellar luminosity and mass, depend upon the 
correct 
determination of their distances.  But the distances to Galactic PNe have 
been problematic for decades, as relatively few individual PN distances are 
accurately determined.  Typically, statistical methods of distance 
determination have been used to derive these important properties, in 
spite of the difficulty of distinguishing nebulae that are optically 
thick from those that are optically thin to H ionizing radiation, where 
different statistical techniques are employed.  
But even if these statistical distances were accurate for a given PN sample, 
it is often hard to draw conclusions when the PN sample so derived is 
small---e.g., when dealing with rare morphologies such as point-symmetric 
PNe. 
Studies of Galactic PNe are also vulnerable to selection biases, such as 
extinction through the Galactic plane that can make certain PN populations 
more difficult to observe.  
For instance, bipolar PNe (and asymmetric PNe in general) are more 
concentrated in the Galactic disk, with a scale height of roughly half 
that of symmetric (round or elliptical) PNe \citep{Manchado_etal00}.  Thus 
asymmetric PNe are somewhat more affected by interstellar extinction, and 
are more likely to fall below detection thresholds.  
In principle, the sample of asymmetric PNe could be 
significantly under-represented, rendering statistical comparisons between 
symmetric and asymmetric PNe less valid. 

The idea of investigating the morphologies of PNe in the Magellanic Clouds 
arises from these difficulties, but it requires {\it HST} imaging to 
determine their classifications with certainty.  {\it HST} observations of 
LMC PNe began with an early program by \citet{Blades_etal92}, and was 
followed up by \citet{Dopita_etal96} and \citet{Vass_etal98}. Unfortunately, 
all of these observations were acquired with the spherically aberrated 
{\it HST} before the first servicing mission in 1993, which limits their 
utility.  \citet{Stang_etal99} summarized these observations 
and expanded the analysis of these data to study the LMC PN morphologies.  
Another, more recent set of LMC PN images were acquired by Dopita ({\it HST} 
Program 6407), but to date they have not been published.  

Our Cycle 8 program (ID: 8271), a snapshot survey using STIS slit-less 
spectroscopy, both dramatically enlarges the existing pool of LMC PN 
images obtained with {\it HST}, and provides the most complete 
information on extra-galactic PN morphology 
achieved to date. This is because our observing technique yields the 
shape of the PNe in several prominent recombination and forbidden lines, 
in addition to the line fluxes, thus giving morphological and spectral 
information at once.  With the fruits of our survey we can use morphology 
as an independent parameter to search for relationships between the physics 
of the nebulae and their progenitor stars, and we can do this better than in 
prior surveys. 
The survey is 58\% complete, but the observing policy for SNAPSHOT 
observations with {\it HST} means that few of the remaining targets are likely 
to be observed. 
If more PNe are observed in the future, we will include and analyze them in subsequent papers. 

While the major motivations for this LMC survey was the detection of PN 
morphology in a sample that was free of distance- and extinction-biases, 
this was not the only motivation.  
The different metallicity of the LMC and the Galaxy allows a comparative 
study of PN formation and evolution 
in environments where the initial (progenitor) chemistry differ.  
An planned extension of our program to the SMC will permit an 
exploration of these effects with an even larger metallicity baseline.  

In this paper we describe our program, from planning to results, 
and we focus on the morphology of the nebulae and their peculiarities.
In $\S$2 we describe our dataset, including the target selection criteria,
the observing techniques, and the calibration process used to obtain the
images of the figures. In $\S$3 we describe our morphological classification, 
the nebular dimensions, and comment on individual characteristics of the 
nebulae. Section~4 discusses the morphology in relation to the evolution of 
the nebulae, explores the variation of nebular expansion velocity with size 
and morphology, and concludes with comments on selection effects and other 
caveats.  Section~5 presents the conclusions and looks forward to future 
projects.

\section {The LMC PN Dataset}

\subsection {Target Selection}

The main tactical thrust of our program is to determine the morphology 
of a large number of PNe in the monochromatic light of a number of important 
emission lines, and to determine the chemical abundances from ground-based 
spectroscopy.  Since {\it HST} images are the most expensive and critical 
commodity, we needed to make maximum use of existing images in the Hubble 
Data Archive, in spite of the poor quality of some of the images (which 
were affected by spherical aberration), and the lack of images in the 
continuum and in low-ionization lines \citep{Stang_etal99}.  We elected 
not to include these targets in our survey, but they may be worth 
re-observing in the future.  

In view of the aim of our scientific project, to correlate morphology with 
the physical and chemical characteristics of the nebulae, their central 
stars, and their immediate environment, we required moderate resolution 
spectroscopy for all PNe in our survey.  For this reason, and to help 
maximize the chance that our selected targets were genuine PNe, we  
selected where possible PNe with published spectra, either from ground-based 
or (ideally) from {\it HST} observations.  We also ensured that our sample 
was widely spread over the face of the LMC in order to explore any 
relationships that might exist between properties of the nebulae and 
location within that galaxy.  
These constraints still left us with a large number of candidate targets, 
so that we were free to exclude PNe in exceptionally crowded fields of 
bright sources: such contamination is particularly problematic for slit-less 
spectroscopy. 
But in the end, the major constraints in our target selection were the 
restrictions that apply to ``snapshot" program with {\it HST}, the most 
severe of which is that they are limited to one orbit per visit; in general, 
the available exposure time is constrained to $\lesssim40$ min, minus 
instrument overheads.  
Within these limits, we tried to select a sample of 50 PNe, out of the few 
hundred PNe known in the LMC \citep{Leisy_etal97}, which spanned as large a 
range as possible in the [\ion{O}{3}] and \ha\ line fluxes in order to 
cover as large a range as possible of nebular age and other properties.  

\subsection {Observations} 

The data presented here were obtained with {\it HST} using the Space 
Telescope Imaging Spectrograph (STIS).  The design for this instrument was 
described by \citet{Woodgate_etal98}, and the initial on-orbit performance 
was summarized by \citet{Kimble_etal98}.  All of our observations were made 
with the CCD detector, using a gain of 1 $e^-$/ADU.  Most of our exposures 
were split into two equal components to facilitate cosmic ray removal.  
We obtained slit-less spectra with the G430M and G750M gratings which yielded 
monochromatic images of the sample nebulae in several important emission 
lines.  
We also obtained direct, broad-band images with the clear (50CCD) aperture 
in order to measure central star magnitudes as faint as $V\approx24$ with 
a short exposure, and to distinguish between the spatial and the velocity 
structure in extended and/or rapidly expanding nebulae.  
The observing log is presented in Table~\ref{ObsLog}.  

The STIS CCD plate scale of 0\farcs051 pixel$^{-1}$ yields a physical scale 
of $\approx0.025$ pc pixel$^{-1}$, which is comparable to ground-based 
images of Galactic PNe.  Only one of the nebulae in our sample had been 
previously resolved from the ground, and few emission line fluxes beyond 
H$\beta$ 4861~\AA\ and [\ion{O}{3}] 4959, 5007~\AA\ had been published.  We 
therefore planned our exposures to achieve a signal-to-noise ratio of 
$\sim30$ per pixel for a typical nebula of 0\farcs5 in 
diameter and a roughly uniform surface brightness distribution, but we 
bracketed our exposure times where possible to avoid saturation (for 
angularly small nebulae) while still providing sufficient signal (for large 
nebulae) to make a good morphological classification.  Finally, the 
limitations of {\it HST} ``snapshot'' programs (e.g., a maximum of one 
orbital period per target) meant that the G430M exposure for the faintest 
target, LMC-J~41, had to be omitted in favor of a strong detection in G750M. 

The G750M spectra cover the wavelength range from 6295--6865~\AA, which 
includes emission from H$\alpha$ 6563~\AA\ and [\ion{N}{2}] 6548, 6583~\AA: 
these are the primary lines used for morphological classification.  For 
those nebulae with relatively high surface brightness, the exposures also 
contain monochromatic images of [\ion{S}{2}] 6717, 6731~\AA, \ion{He}{1} 
6678~\AA, and (for low-ionization nebulae) [\ion{O}{1}] 6300, 6363~\AA.  
Together, these lines reveal the low- to moderate-ionization morphological 
features, which are the most important for tracing the earliest phase of the 
morphological evolution of the nebulae.  The mean dispersion for G750M is 
0.56 \AA\ pixel$^{-1}$ (26~km s$^{-1}$ pixel$^{-1}$), so that no overlap 
of the monochromatic images occurred in the [\ion{S}{2}] lines, nor between 
[\ion{N}{2}] 6548~\AA\ and H$\alpha$, for nebulae less than 
$\approx$1\farcs4 in diameter.  Some of the nebulae are larger than this, 
and for these cases the continuum images were used to help distinguish the 
spatial features from velocity structure.  (The overlapping lines for a 
few of the nebulae do not affect the measurements and classifications 
presented in this paper.) 
The G430M spectra cover the wavelength range from 4818--5104~\AA, and 
include emission from H$\beta$ 4861~\AA\ and [\ion{O}{3}] 4959, 5007~\AA. 
These lines provide the morphology at intermediate ionization, and can also 
be used to infer the ionization structure within the nebulae.  The mean 
dispersion of 0.28 \AA\ pixel$^{-1}$ (17~km s$^{-1}$ pixel$^{-1}$) means that 
the internal velocity structure of some of the nebulae could have been 
resolved, though this did not impede our ability to assign a morphological 
classification. 

\subsection {Calibration} 

The images and slit-less spectrograms were calibrated with the standard 
STIS calibration pipeline, version 2.3, released 1999 October 
\citep[see][]{Hodge_etal98b}.
Processing in all cases included basic 
two-dimensional reductions, as described in \citet{Hodge_etal98a}, including 
corrections for bias and dark current, flat-fielding, and the combination 
of the paired exposures to remove cosmic rays.  Special care was taken with 
the correction for the dark current because the temporal variation, 
which is related to the constant exposure of the CCD to high energy charged 
particles, is quite significant on a timescale of days.  The calibration 
reference files for the dark correction are available from the {\it HST} 
archive approximately monthly, although individual dark frames are obtained 
on a daily basis.  We down-loaded a script from the ST~ScI Web site which 
enabled us to combine the so-called ``weekly'' dark reference files with 
$\sim5$ individual dark frames that were obtained within a few days of each 
observation.  The script corrects new hot pixels that deviate by more than 
5-sigma from the weekly dark by using the median of the daily dark frames 
at that location.  The resulting images are significantly improved compared 
to those processed without the constructed dark frame.  However, it is 
apparent that a number of hot pixels are not corrected, or are not 
adequately corrected, by this procedure, which compromises the photometric 
accuracy on small spatial scales.  Fortunately, none of the measurements 
or assessments offered in this paper are affected in a significant way by 
the imperfect dark correction. 

\section {Dimensions and Morphology}

The broad-band and [\ion{O}{3}] 5007\AA\ images for the target nebulae 
(apart from three: see \S~3.1 below) are presented in Figures~\ref{Clear_1} 
through \ref{Clear_11}; the broad-band data are rendered as grey-scale 
images, and the [\ion{O}{3}] images are rendered as contour plots.  
The grey-scale mapping is either the log or the square-root of the image 
intensity, in order to bring out the often faint structural features.  
The plots show contours on a linear scale, with 10\% intensity 
intervals, to illustrate the structural features that are most relevant for 
the morphological classification.  

We classified the morphologies in our sample from the [\ion{O}{3}] 5007 \AA\ 
monochromatic images in the G430M spectra (although we were guided by the 
\ha\ and [\ion{N}{2}] 6548 \AA\ images), using a classification scheme 
similar to \citet{Manchado_etal96}.  This classification scheme was recently 
extended \citep{Gorny_etal97, Stang_etal99} to recognize explicitly the 
bipolar core (BC) nebulae that show distinct hemispheres of emission in 
the nebular core. 
Usually, but not always, such nebulae were classified as ``Rs'' or (more 
often) ``Es'' in the \citet{Manchado_etal96} scheme, meaning that such 
nebulae have round or elliptical outer contours, but have internal 
structure.  Our designation of BC indicates the presence of a bi-nebulous 
structure with an intensity contrast of $\gtrsim20$\%.  When present, 
such structure is in our view the more important morphological feature, 
in that these nebulae are more closely connected to pure bipolar (B) PNe 
than either R or E (see \S~4, below).  Support for this view also comes 
from L.~Stanghellini (2000, in preparation), who found that the BC PNe in 
the Galaxy have a disk scale height that is much more similar to that 
of B than to E or R nebulae.  
Our distinction between E and R was based on whether the major axis of the 
10\% intensity contour exceeded the minor axis by more than 10\%.  

Usually the morphological class we assigned on the basis of structures evident 
in the [\ion{O}{3}] image would not have been different had we instead used 
another, lower-ionization line such as [\ion{N}{2}] 6583 \AA.  But in four 
cases (15\% of the resolved nebulae) the classification might have changed: 
we identify these cases in the notes on individual nebulae, below.  
Table~\ref{Morph} gives our classifications and the nebular dimensions. 
The nebular dimensions were measured with respect to the 10\% intensity 
contour of the outermost structure, and are presented in 
column 5 of the table.  We also measured the formal photometric radii, 
according to the method described by \citet{Stang_etal99}.  This measurement 
gives an objective measurement of nebular angular size which is insensitive 
to the S/N ratio of the image, and is useful for 
evolution studies.  It corresponds to the size of a circular aperture that 
contains 85\% of the flux in [\ion{O}{3}] 5007 \AA, and is given in 
column 4 of the table.  

\subsection {Individual Nebulae}

We describe in this section the morphological details for each nebula 
listed in Table~\ref{Morph}. 

{\bf J~41:} This nebula has two distinct shells with elliptical outer 
contours, and an inner core that is bipolar in the light of \ha. 
The classification was based solely on the \ha\ image, since 
no other emission lines were available.  The central star is easily detected. 

{\bf SMP~4:} This elliptical PN has a faint, elliptical outer shell: it 
could be classified as E/AH following the \citet{StangPasq95} classification 
scheme for multiple shell PNe.  The central star is easily visible in the 
continuum image.  

{\bf SMP~9:} This nebula has a barrel shape, but the outer contour is elliptical.  In the emission line images there is a marginal detection 
of a central equatorial enhancement (i.e., a ring).  
No central star is detected. 

{\bf SMP~10:} This spiral-shaped object has a classic point-symmetric 
morphology that is strikingly like that of a spiral galaxy.  Although this 
morphology is unusual for a PN (and it is the only such PN known in the LMC), 
there is a Galactic PN (``GPN''), K4--55 \citep{Manchado_etal96} that looks 
very similar. 
SMP~10 could also be an {\it incomplete bipolar}, seen face on (see NGC~6309 
in the catalogs by Balick 1987 and Schwarz, Corradi, \& Melnick 1992).  
In SMP~10 two limbs, or arms, project obliquely on opposite side of an 
elliptical, almost round main body: one east of north, 
and the other west of south. In the [\ion{N}{2}] images they measure about 
0\farcs1 at maximum thickness and 1\farcs3 in extent. 
The nebula is of moderate overall ionization, but the {\it arms} include 
a region of distinctly lower ionization: these features are readily 
apparent on the [\ion{N}{2}] 6548,83 \AA\ lines but are less distinct in 
the \ha\ image, and even less so in the [\ion{O}{3}] images.  The presence 
of these features is reminiscent of the Galactic PN NGC~6543, though of 
course the spatial resolution is insufficient to determine whether a jet 
from the central star is present and related to the low ionization emission 
regions. 
The central star is easily visible in the continuum image.  

{\bf SMP~13:} This nebula has a round outer contour, with a bipolar, if 
somewhat patchy, inner core.  
There is little if any stratification in the ionization.  
The central star is easily detected. 

{\bf SMP~16:} This is a bipolar planetary nebula with thick ring and a classic 
``butterfly'' shape.  Knots of emission are apparent along the ring in the 
[\ion{N}{2}] images, but are less distinct in the \ha\ and [\ion{O}{3}] images. 
The ionization varies throughout the nebula: it is relatively high in the 
ring and the ``wings'' of the butterfly shape, but drops considerably near 
the outer edges of the wings.  The nebula resembles the GPN NGC~2346. 
The central star is not visible through the ring. 

{\bf SMP~18:} The uncertain classification of this PN is a result of the 
contrast between its elliptical (possibly with a multiple shell) appearance 
in the continuum image, and either a slight asymmetry or a small central 
cavity the \ha\ and [\ion{O}{3}] images.  
The central star may be marginally detected. 

{\bf SMP~19:} This PN has an elliptical outer contour, but a clearly 
bipolar core.  The ring-like structure is most apparent in the [\ion{N}{2}] 
images.  The central star is marginally detected. 

{\bf SMP~25:} This small PN has an elliptical outer contour, and a Gaussian 
intensity profile.  The central star is easily detected. 

{\bf SMP~27:} The core of this object has the same quadrupolar morphology 
in all detected lines, including the broad-band image.  Two extended nebular 
features are evident, though faint, in H$\alpha$, [\ion{O}{3}] 5007, and 
the broad-band image.  The outermost is an arc approximately 6\farcs25 
to the NW, which is directed inward (to the CS), with a curvature that 
is somewhat larger than that of a circle centered on the CS.  This arc is 
approximately 7\farcs5 long and 0\farcs5 wide, and has a surface brightness 
$\sim10$ times fainter than the central nebula.  The other feature is a 
blob centered approximately 4\farcs8 to the north, with dimensions 
2\farcs7 x 1\farcs0, with a similar surface brightness to the arc.  The 
blob is roughly aligned with the major axis of symmetry of the inner nebula. 

The central (bright) region of this object is very highly ionized: there is 
no detection of [\ion{S}{2}], [\ion{O}{1}], and only very weak [\ion{N}{2}].  
On the other hand, [\ion{O}{3}] 5007, 4959 are very strong relative to 
H$\beta$, and \ion{He}{1}~6678 is easily detected.  A ratio of the 
[\ion{O}{3}] 5007 to H$\alpha$ images shows a relatively uniform inner 
structure, though the ratio falls to roughly one-third of the core value in 
the outer structures.  Though the ionization apparently declines with 
distance from the CS, the ratio could conceivably be affected by a higher 
electron temperature or, alternatively, a higher O$^{++}$ abundance in 
the inner region; long-slit spectroscopy with STIS can resolve this 
question. 

It is very tempting to associate the outer features with the central PN, 
especially since it is otherwise not obvious how this gas is ionized. 
The features are not quite circularly symmetric, though this is not unusual 
for GPNe.  Also, the outer features are at the limit of detection 
on these images, and other features may lurk beneath the noise.  
The central star is bright, and easily detected in the continuum image and 
in the dispersed images. 
SMP~27 has a measured expansion velocity of 33 km s$^{-1}$ 
\citep{Dopita_etal88}.  If this velocity 
were constant throughout the life (and spatial extent) of the nebula, the 
implied dynamical age is 3500 yr for the inner region, and $\sim45,400$ yr 
for the outer arc.  A plausible explanation for the presence of the outer 
ionized structure is that the ionization front from the inner nebula has 
encountered remnant gas from the AGB wind that preceded the formation of 
the inner nebula, and the subsequent evolution of the CS to high temperature. 
The near-alignment of the ionized blob and the symmetry axis of the inner 
nebula suggests a kinematic connection to an asymmetric ejection from the CS. 

{\bf SMP~28:} At the first sight, this PN can be considered {\it bi-nebulous} 
\citep{Greig72}, in that its main shape comprises two distinct blobs.  
Closer inspection of the \ha\ and [\ion{N}{2}] images reveals three roughly 
co-aligned blobs, separated by 0\farcs2.  The central blob is the brightest, 
and corresponds to the position of the central star, which is apparent on 
the continuum image.  Two faint, thin arms are evident in the 
light of [\ion{N}{2}] that extend obliquely from opposite sides of the 
central blob, which may be similar to the arms seen in SMP~10. 

{\bf SMP~30:} This object is morphologically similar to NGC~6853 in the 
Galaxy.  Although this PN is 
clearly asymmetric, it is difficult to say for certain whether the morphology 
is bipolar or quadrupolar.  The ionization structure is relatively complex: 
the ionization in general declines from center to the periphery, but there 
are tight knots of [\ion{N}{2}] bright material at the ends of an equatorial 
axis that are rather more diffuse in the light of [\ion{O}{3}].  
A very faint star is detected close to the center of symmetry, which is 
probably the central star. 

{\bf SMP~31:} This PN is barely resolved; the classification as ``R" is 
therefore uncertain.  No central star is apparent. 

{\bf SMP~34:} Although this PN appears round in the continuum image as well as 
the \ha\ and [\ion{O}{3}] images; there is a hint of bipolar structure in 
the light of [\ion{N}{2}].  The central star is detected. 

{\bf SMP~46:} This PN is ring-like in appearance, with a bright knot on 
the southwestern edge.  No central star is detected. 

{\bf SMP~53:} This elliptical, barrel shaped PN does not clearly show a 
bipolar core.  Nonetheless, a ring of emission is very clear in the light 
of a [\ion{N}{2}] (and to some extent in \ha), thus we classify it BC. 
No central star is detected. 

{\bf SMP~58:} This PN is barely resolved; the classification as ``R" is 
therefore uncertain.  The central star is probably detected.  
We do not show this object in the figures, since there are no significant 
morphological features or nearby field stars of comparable brightness. 

{\bf SMP~59:} This object has quadrupolar morphology in all detected lines, 
including the broadband image.  It shows a lot of internal structure, 
with a very knotty major axis, and a fainter minor axis that is not orthogonal 
(in the plane of the sky) to the major axis.  The ionization structure 
generally follows the spatial structure, except that the knots of emission 
are spatially more confined in the light of [\ion{N}{2}].  The bright central 
star is easily detected in both the continuum and the dispersed images. 

{\bf SMP~65:} This PN would be a classic round PN were it not that the core 
of the \ha\ + [\ion{O}{3}] emission is slightly displaced $\sim$0\farcs1 
north of the CS.  This object also has a faint, outer halo.  The ionization 
is spatially fairly uniform, and relatively high, with no [\ion{N}{2}] or 
other low-ionization species present in the spectrum.  
No central star is detected. 

{\bf SMP~71:} This PN is elliptical in every contour of constant intensity, 
but the major axis of the ellipse in the inner (bright) region is 
orthogonal to that of the outermost contour.  
The central star is not detected. 

{\bf SMP~78:} This object shows a barrel-shaped morphology very similar 
to that of SMP~53, except that the bipolar core is more evident. 
There is little ionization stratification, and no central star is detected. 

{\bf SMP~79:} This PN has an elliptical outer contour, with a bipolar 
core.  This is the physically smallest BC nebula in the sample, with a 
radius of 0.05 pc, and a dynamical age of $\sim1400 $yr, assuming a constant 
expansion velocity of 37 km s$^{-1}$ \citep{Dopita_etal88}.  
The central star is not detected. 

{\bf SMP~80:} This is a round PN in continuum light, and in the light of \ha\ 
and in the [\ion{O}{3}] lines.  However, the [\ion{N}{2}] lines reveal an 
inner ring or bipolar shell that is $\approx$0\farcs4 in diameter. 

{\bf SMP~81:} This PN is barely resolved; the classification as ``R" is 
therefore uncertain.  The central star is detected. 
We do not show this object in the figures, since there are no significant 
morphological features or nearby field stars of comparable brightness. 

{\bf SMP~93:} This object is perhaps the most visually striking of the PN 
images presented here; we refer to it as the ``monkey head'' nebula.  The 
morphology in continuum light and in all the detected emission lines shows 
a set of at least three asymmetric, interconnected rings.  The strong 
[\ion{N}{2}] lines show faint emission ``bubbles'' beyond the outermost 
rings, along the long axis of symmetry.  With a maximum extent of 6\farcs4, 
or 1.57 pc, this object ranks with the very largest PNe known in the Galaxy. 
The morphology is closely analogous to the GPN M2--53.  
No central star is apparent. 

{\bf SMP~94:} This object is not resolved (and hence is not shown in the 
figures).  The dispersed images show a double-peaked velocity structure, 
with a width of $\sim$200 km/s.  The ionization is fairly high, but peculiar, 
with prominent \ha, rather weak [\ion{O}{3}] lines, and broad \ion{He}{2} 
6830 \AA\ emission as well.  These features suggests that this object may 
not be a PN, but could be a symbiotic star, as suggested by 
\cite{Dopita_etal88}.

{\bf SMP~95:} This PN has an elliptical outer contour, but a clearly 
bipolar core, with faint ansae extending from the ends of the major axis 
of symmetry.  This object shows moderate ionization, with prominent 
[\ion{O}{3}] lines, but relatively strong [\ion{O}{1}], [\ion{N}{2}], and 
[\ion{S}{2}] lines as well.  No central star is detected. 

{\bf SMP~100:} This is a most interesting PN, with rather complex 
morphological features and ionization structure.  This object has 
an elliptical outer contour and a bipolar (or possibly quadrupolar) core 
denoted by bright lobes at either end of the minor axis, though the lobe 
to the NW is brighter by nearly half and much more 
spatially extended.  The structure in the [\ion{O}{3}] lines is 
similar to that in \ha, except that the brightness contrast in the lobes is 
much higher in [\ion{O}{3}].  The low-ionization [\ion{N}{2}] lines, on the 
other hand, are only detected in knots that define a ring, with the strongest 
knots lying just outside the ends of the [\ion{O}{3}] emission lobes.  
The central star is somewhat faint, but is easily detected. 

{\bf SMP~102:} This object is a round PN, with many emission knots 
surrounding a central cavity; it may have a bipolar core.  
The ionization is relatively high, with no emission detected in the 
[\ion{N}{2}] lines {\it except} in the bright knot lying to the SW of the 
moderately bright central star. 

\section {Discussion}

\subsection {The Full Sample}

With this dataset, together with the 31 resolved LMC PNe for which 
monochromatic images exist in the {\it HST} archive, we can for the first 
time compare the incidence of PNe among morphological types between the 
LMC and the Galaxy, and search for evidence of temporal evolution among 
morphological types. 
For this purpose we adopted the dimensions and morphological data from 
\cite{Stang_etal99}.  We also retrieved archival {\it HST}/WFPC2 images of 
14 additional LMC PNe from the GO program 6407 (PI: M. Dopita).  These 
images were obtained with narrow-band filters centered on [\ion{O}{3}] 
5007 \AA\ and \ha\ 6563 \AA\ (though in the latter case there is some 
contamination from [\ion{N}{2}] 6548 \AA, which is of no consequence here). 
Our dimensional measurements and morphological classifications, which were 
performed exactly as for the STIS data presented above, are independent of 
the quality of the calibration.  However, the derivation of the mean surface 
brightness (discussed below) does require a good calibration, so we 
re-calibrated the WFPC2 data using the standard pipeline, and performed 
cosmic-ray rejection when more than one frame was available.  The 
dimensions and our morphological classifications for these objects are 
given in Table~\ref{Dopita}. 

\subsection {Morphology and Nebular Evolution}

In Table~\ref{Mtypes} we compare the relative number of PNe in each  
morphological class from the combined LMC sample with that in the Galaxy, 
based on our classification of the PNe in the sample defined by 
\cite{Manchado_etal00}.  
We re-classified all of these nebulae because the BC class was not 
recognized as important in itself when these catalogs were published, and 
because we were interested to see whether our classifications agreed those 
of other experts in the field.  
Our BC nebulae corresponded to a subset of what had been classified as 
E or R in the earlier catalogs---and very often these nebulae were given 
a sub-class of ``s'' to denote some degree of apparent structure.  Apart 
from this distinction the disagreements were rare. 
For the comparison between the LMC and the Galaxy, the B, BC, and Q, types 
were combined to form a single ``asymmetric'' class of nebulae.  
This comparison is important because bipolarity has been shown to be 
an indication of Population~I ancestry for PNe in the Galaxy, and recently 
this relationship has been extended to the LMC by \cite{Stang_etal00}. 
Though the number of classified nebulae in the LMC (59) is $\sim1/5$ the 
number in the Galactic sample, the fractions of elliptical vs. asymmetric 
nebulae are quite different between the LMC and the Galaxy.  
Taken at face value, this result would imply that a much larger fraction of 
the PNe in the LMC were produced by a younger Population of stars than in 
the Galaxy.  If so, the most direct interpretation would be that the star 
formation history of the LMC PNe progenitors is very different than 
that of the Galaxy (averaged over the volume defined by the Galactic PN 
sample).  While there is evidence to support a burst of star formation 
occurring in the LMC within the last 4 Gyr \citep{Grebel98}, it may well 
be that some, or perhaps most, of the discrepancy between PN morphological 
classes can be attributed to selection effects in either PN sample.  

\begin {figure}
\plotone {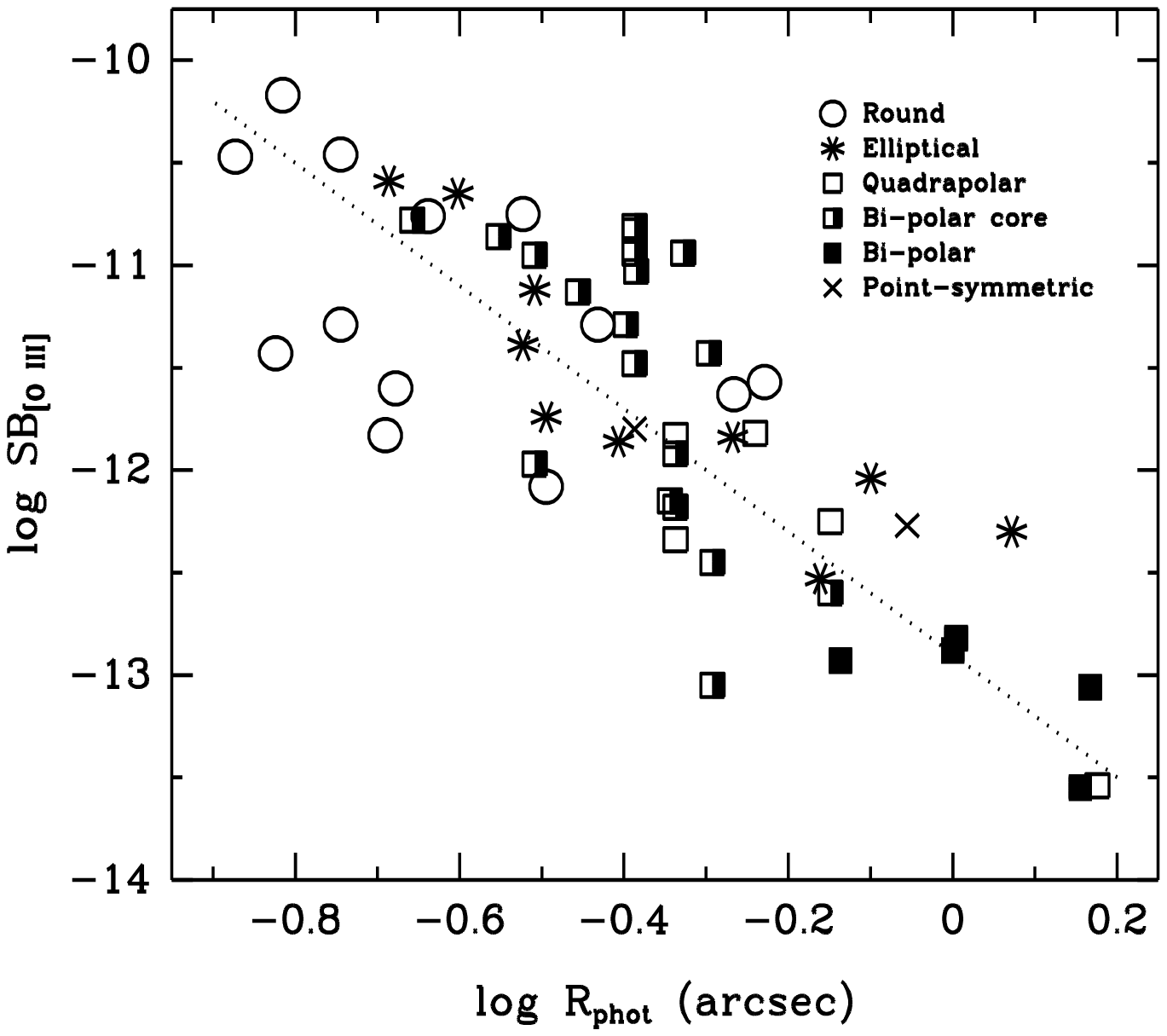}
\caption 
{The decline of surface brightness, in the light of [\ion{O}{3}] 5007\AA, 
with nebular (photometric) radius is consistent with an $R^{-3}$ power law 
({\it dotted line}).  The various morphological types are represented by 
different symbols, as shown in the legend. 
\label{SB_rad}}
\end {figure}

We next explore the question of temporal evolution of nebular morphology 
by examining it in the context of the change in nebular surface brightness 
with nebular size.  All of the nebulae are resolved, though three 
are smaller than 0\farcs30 (6 pixels) in diameter.  
We present the relationship in Figure~\ref{SB_rad}, where the various 
morphological types are mapped to symbols according to the legend.  For 
this purpose, we actually derive the mean surface brightness in the 
[\ion{O}{3}] 5007\AA\ line within the photometric radius, corrected for 
interstellar extinction according to the formula: 
\begin{displaymath}
	I_{corr} =  I_{obs} \times 10^{c*(1+f_{\lambda})}
\end{displaymath}
where $c$, the logarithmic extinction at \hb, is taken from 
\citet{MD91a, MD91b} and $f_{\lambda}$, the extinction law from 
\citet{Howarth83} for the LMC, is approximately 0.02 at 5007\AA.  
Since this sample is drawn entirely from the LMC, the radius is linearly 
proportional to the physical size of the nebula, which is also a rough 
indicator of nebular age (but see \S~4.3 below).  
The dotted line in the figure corresponds to a decline in nebular 
surface brightness proportional to $R^{-3}$, which is roughly consistent 
with the data.  The actual change in 
surface brightness of an evolving PN must relate in a fairly complex 
way to the expansion of the nebula, the change in ionization, temperature 
and density of the gas, and the simultaneous and potentially very large 
changes in the temperature and luminosity of the central star as it evolves 
during the PN lifetime.  That this relationship could be so simply 
characterized (albeit with large deviations) is interesting, and may help 
constrain evolutionary models. 

The figure shows that ten of the eleven angularly smallest nebulae 
($0\farcs13\leq R_{Phot}\leq 0\farcs26$) are round or elliptical.  In this 
size regime most of the nebulae are well resolved (i.e., exceed in size 
a STIS CCD spatial resolution element of 2 pixels), 
and the detected signal per pixel is very high (usually $>10^4$ counts 
pixel$^{-1}$, and $>10^5$ counts in total).  However, three of the nebulae 
have radii of $\le0$\farcs15; the morphological classification in these 
cases is less certain because faint structural details may not be resolved.  
The lack of asymmetric nebulae at small sizes could be interpreted as an 
evolutionary effect: that is, any initial asymmetry in the gas distribution 
and velocity field of R or E nebulae may not have had time to manifest itself 
in the morphology, while extreme B nebulae (the progeny of the most 
massive and luminous central stars) may evolve so quickly through this 
phase that they are unlikely to be detected.  As we will show in the next 
subsection, the asymmetric nebulae in this sample have higher mean 
expansion velocities. 
But we cannot rule out the possibility that observational selection is 
largely responsible for the disproportionate fraction of R+E nebulae at 
small size, mostly because the catalogs from which we selected our sample 
contain PNe that are relatively bright in [\ion{O}{3}] 4959, 5007\AA, the 
presence of which is often used as a criterion for classification as a PN. 
Certainly the small number statistics in this size regime 
makes firm conclusions about evolutionary effects for small PNe difficult.  
In any case, the onset of asymmetric features is known to appear in very 
young Galactic PNe (e.g., M~2-9 and CRL~2688), and even SMP~74 in the LMC 
is probably younger than $2000$~yr.  
This suggests that the gross features of the nebular morphology are well 
connected to PN formation.  
What is more clear in our sample is the segregation of bipolar core PNe 
(between $0\farcs30\leq R_{Phot}\leq 0\farcs70$), which have a round or 
elliptical outer contour, and pure bipolar nebulae ($>0\farcs70$, or 
$>0.18$~pc). 
This suggests that the bi-polarity may become the dominant morphological 
feature during the lifetime of BC nebulae, perhaps 
through subsequent shaping by the radiation field and wind from the 
central star \citep{Balick87}.  This result, when combined with the 
identification of BC with B as the products of a younger stellar population 
\citep{Stang_etal00}, lends support to the identification of the BC class 
as a physically important morphological feature.  

\subsection {Nebular Expansion}

\begin {figure}
\plotone {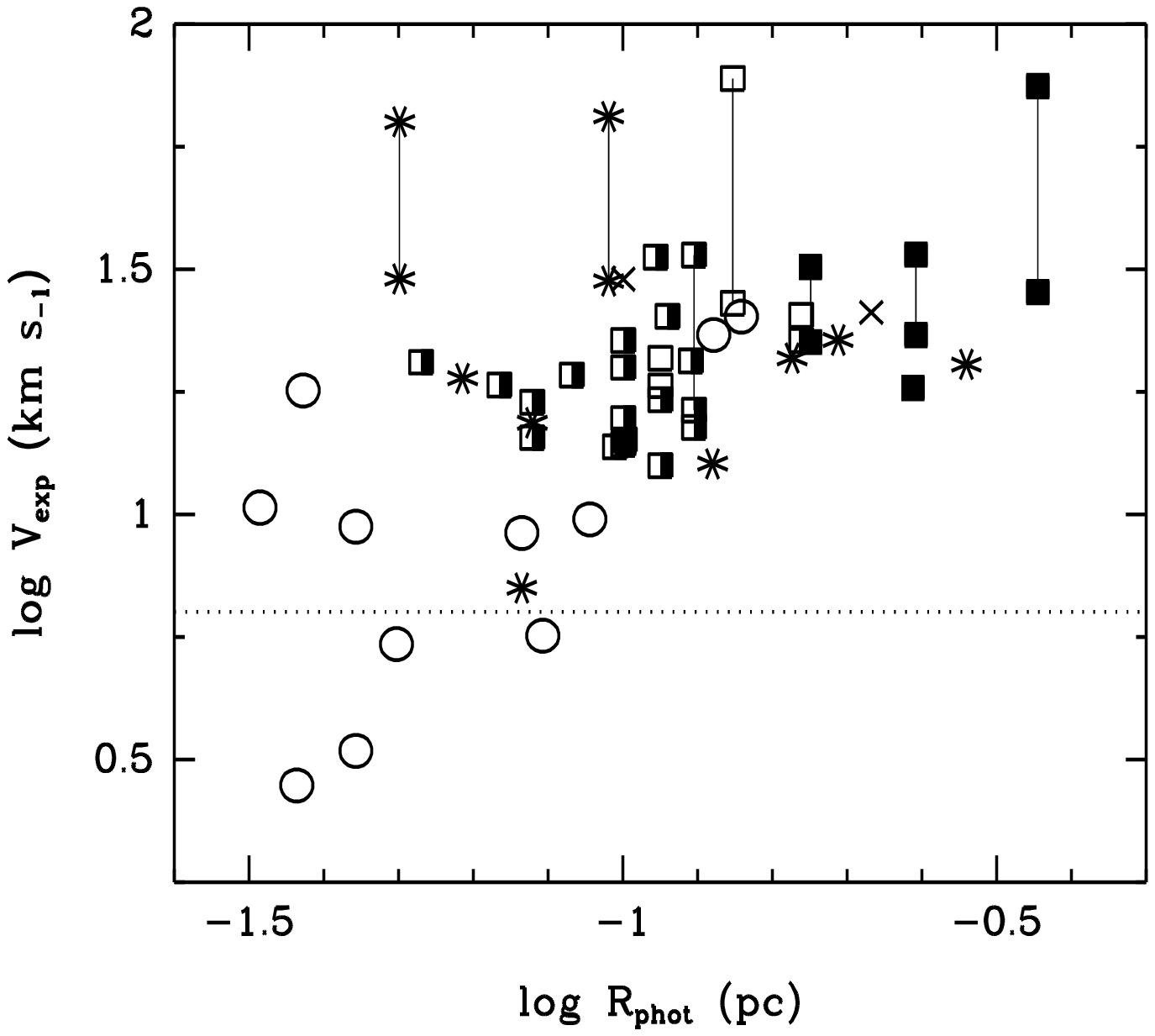}
\caption 
{Nebular expansion velocity \citep[adapted from][see text]{Dopita_etal88} 
as a function of nebular physical radius.  Greater uncertainty likely 
applies to values of V$_{exp}$ for which the correction for the 
instrumental resolution ({\it dotted line}) is large.  Nebulae in which 
multiple expansion components were detected are shown at the extrema of 
the published velocity range, and are connected with vertical lines. 
Symbols as in Fig ~\ref{SB_rad}. 
\label{V_exp}}
\end {figure}

We alluded above to the connection between the physical size of the PN 
and its age, i.e., the time elapsed since the visible PN shell was 
ejected from its progenitor.  The ``dynamical" age, the ratio of the 
physical size to the current expansion velocity, has been used for this 
purpose under the assumption 
that the expansion has been constant throughout the life of the nebula.  
The validity of this assumption has been questioned by many, and 
some \citep[e.g.][]{Dopita_etal87, DopitaMeth90, Dopita_etal96, Vass_etal98} 
have argued that the nebular shell 
experiences acceleration during the transition of the CS to high 
temperature at constant luminosity.  We explore this point in 
Fig.~\ref{V_exp}, where we show the variation of nebular expansion 
velocity with physical radius for this LMC sample.  The physical radius 
was derived from our data by scaling the photometric radius by 
0.245 pc arcsec$^{-1}$; 
the velocities, derived from the [\ion{O}{3}] 5007~\AA\ line, were taken 
from \citet{Dopita_etal88} and corrected according to \citet{Stang_etal99} 
(i.e., those derived by measuring the width of single line profile were 
reduced by a factor of 1.82).  
The most striking feature of this plot is that, for nebulae smaller than 
$\sim0.1$ pc, the expansion velocities deviate, and are generally less 
than, the mean of $\sim20$~km~s$^{-1}$.  The significant differences 
between this plot and that in \citet[][Fig.~4]{DopitaMeth90} are that the 
nebular radii are much more accurate, and that the morphological types 
are known.  
All of the PNe in this range are of type R or E, though the types of the 
three smallest nebulae are less certain (see \S~4.2).  
A possible interpretation is that nebular shells experience an acceleration 
from the central star until they reach a critical size, by which time the 
acceleration becomes negligible either because of geometrical dilution of 
the shell or because the central star luminosity has substantially declined. 
Which effect applies would depend upon the relative rate of evolution of 
the shell vs. the central star: the less massive CSs have lower maximum 
luminosities, and therefore might be expected to provide less acceleration.  
At the other extreme, another possibility is that acceleration of the 
shells is nearly zero in these PNe (i.e., those larger than 0.03 pc), but 
that some nebulae (mostly of type R or E, evidently) expand very slowly.  
In this case, the nebulae would have to fade below the detection limit 
before they reach $\sim0$\farcs1 in size.  

We believe it is difficult to draw firm conclusions about shell 
accelerations, and therefore about nebular dynamical ages, based upon the 
available velocity data.  The spectra of the [\ion{O}{3}] 5007~\AA\ 
profiles were obtained at $\approx12$ km~s$^{-1}$ resolution 
\citep{Dopita_etal88}, so that in a number of cases the correction for 
the instrumental profile was comparable to or exceeded the measurement 
(see the dotted line in Fig.~\ref{V_exp}).  Ignoring these points 
significantly weakens the case for acceleration in this size regime.  
More importantly, the spectrograms did not resolve the nebulae spatially, 
so that high velocity components, when detected, cannot be associated 
directly with common morphological details seen in these or other nebulae, 
such as shells, rings, FLIERS, or jets.  That is, at least for some nebulae, 
it is not clear whether 
the expansion velocity measured from this one moderate-ionization line is 
representative of the nebular expansion as a whole.  To illustrate the 
point, nebulae in which multiple velocity components were detected are 
connected in the Figure with symbols at the extrema of the velocity range.  
Detailed interpretation of complex velocity structures in PNe, such 
as that published by \citet{Guerrero_etal00} for NGC~6891, is very 
important for establishing the kinematics of the expanding nebula.  
We believe it is important to obtain spatially resolved, high resolution 
spectroscopy of the LMC PNe to relate reliably the expansion velocities 
to nebular dynamical ages.  

\subsection {Caveats and Selection Effects}

Although we believe the morphology of PNe in the LMC reveals an important 
clue about the Population of the progenitors, we must be mindful of selection 
effects that may affect our interpretation.  As we described in \S~2.1, our 
sample was drawn from surveys that are known to be incomplete, either in 
depth or in spatial coverage \citep{Jacoby80, Leisy_etal97}.  Since few 
known LMC PNe are resolvable from the ground, the spectroscopic criteria 
for identification as PNe often includes bright [\ion{O}{3}] emission, 
which selects against the youngest PNe (where the ionization has not yet 
reached the point where significant O$^{+2}$ has been produced), and the 
very oldest PNe (where the ionization has declined because of the dimming 
and dilution of the CS radiation field).  Few of these surveys probe very 
deeply, and in any case our program did not include 
more than a very few PNe that were moderately faint (in either or both of 
\ha\ and [\ion{O}{3}]).  Figure~\ref{SB_rad} shows that, to first order, 
excluding the faintest PNe also excludes the dynamically oldest PNe.  
The Galactic sample may also suffer from selection bias.  For instance, the 
asymmetric PNe may be under-represented in the Galactic sample since they 
are more closely confined to the disk, and hence suffer more attenuation 
(on average) from interstellar dust, which would select against their 
discovery. 

\begin {figure}
\plotone {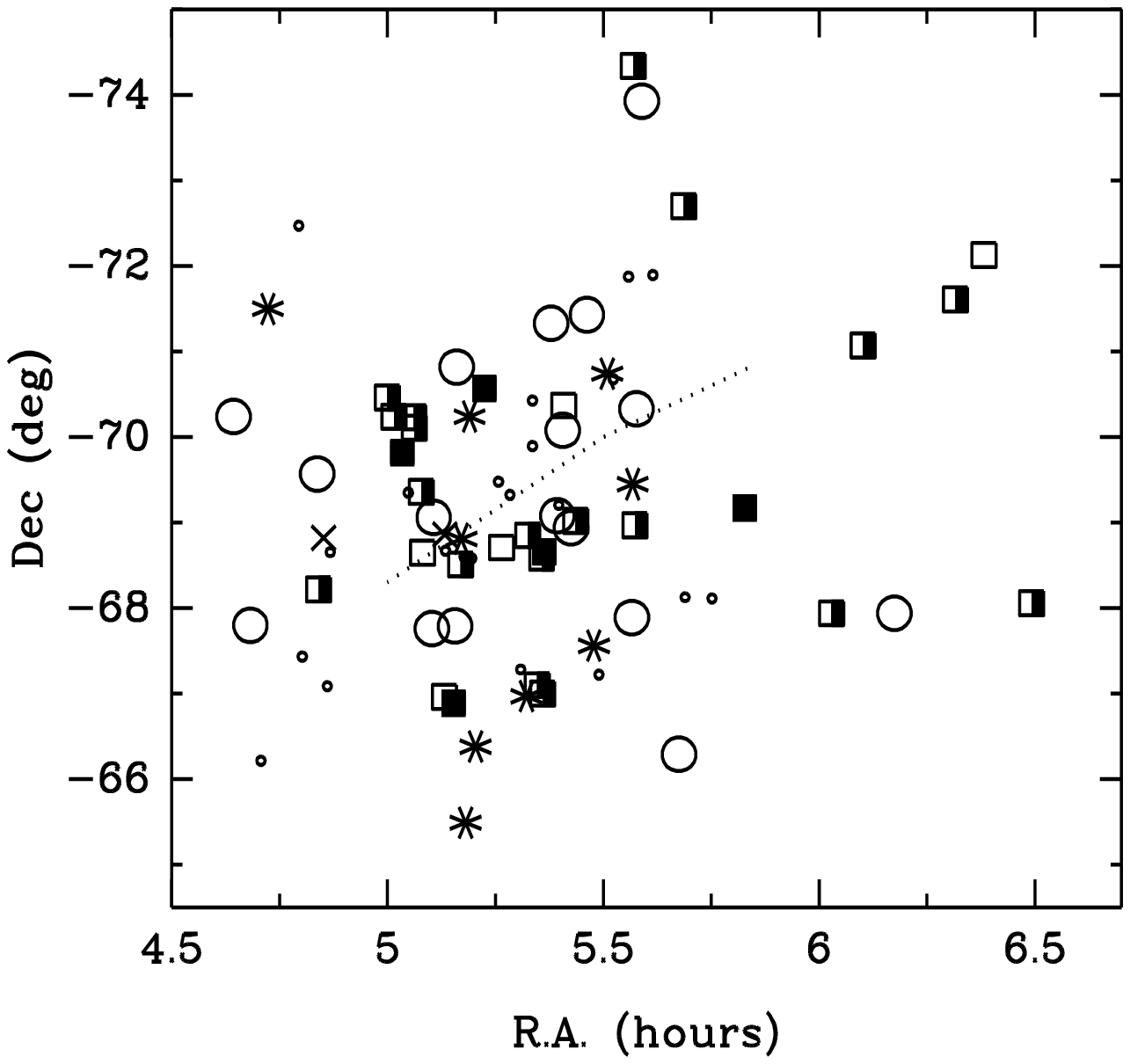}
\caption 
{Surface distribution of PNe in the LMC that have been imaged with 
{\it HST}.  Symbols as in Fig ~\ref{SB_rad}, except that small circles 
denote PNe in our SNAP program that have not yet been observed.  The 
approximate location of the axis of the LMC bar is shown 
({\it dotted line}).
\label{RA_Dec}}
\end {figure}

Finally, the spatial coverage of the discovery surveys does not extend 
to the faintest 
regions of the LMC \citep{Morgan94}, which excludes PNe that could be 
the progeny of a much older population of stars.  Fig.~\ref{RA_Dec} shows 
the spatial distribution of this sample of PNe, relative to the LMC bar, 
coded by morphological type.  Most of these PNe lie within $\sim$2\degr\ of 
the bar, which is consistent with the optical continuum distribution of 
light from stars.  We know of no particular segregation of population 
types (other than young stars in \ion{H}{2} regions) in the LMC, though 
some studies suggest that tidal interactions with the Galaxy greatly 
perturb the orbits of moderate to old stars (e.g., Weinberg 2000).  
Certainly the spatial distribution of nebular morphologies in 
Fig.~\ref{RA_Dec} shows no particular segregation within the LMC, other 
than that there are more PNe of all types in the bar.  While this 
distribution is consistent with that for the older stellar Populations, 
it is probably fortuitous given the selection effects mentioned above.

\section {Conclusion and Future Work}

We have used the {\it HST} and STIS in slit-less mode in a campaign to obtain 
monochromatic images of a large sample of Magellanic Cloud PNe in several 
important emission lines.  The set of LMC PNe presented here display a wide 
variety of morphological features, including examples of the rare Quadrupolar 
and Point-Symmetric classes.  We argue that bipolar core PNe are 
a variant of, and sometimes an evolutionary precursor to, 
the pure bipolar class.  Such a finding is consistent with published 
evidence that BC nebulae are closely related to type B PNe (and are not 
similar to E or R PNe) in chemical abundances, progenitor Population type, 
and (for Galactic PNe) disk scale height.  
It is important that future studies compare the relative ages of BC and B 
nebulae through other means, such as the evolutionary state of the central 
stars, to confirm our suggestion that some or most BC nebulae evolve to B.  
It is also very important to increase the sample size of observed B and BC 
nebulae in the Magellanic Clouds, and to obtain high resolution, spatially 
resolved spectrograms to interpret properly the kinematics of the expanding 
shell(s), and therefore the nebular dynamical ages.  

The nebulae presented here roughly double the number of LMC PNe that have 
been imaged with {\it HST}, and in that combined dataset we find a 
larger fraction of asymmetric PNe in the LMC than in the Galaxy.  
This result either suggests a difference in the stellar Population types 
and/or star formation history between the LMC and the Galaxy, or it shows 
us the extent to which selection effects operate in studies of PN morphology.  
We also find evidence for evolution in nebular morphology, but the extent to 
which it is determined by PN formation processes vs. subsequent star + wind 
interactions is less clear: this question must be pursued by including 
younger PNe in the study.  In any event, the smallest (and presumably the 
youngest) BC nebula in our sample is $\approx 0.05$ pc in radius, implying 
a rough dynamical age of $\sim1400$ yr, which shows that the onset of 
asymmetrical 
features can occur very early in the PN lifetime.  The implication is that, 
for at least some nebulae, the gross morphological features are more closely 
tied to PN formation, and that subsequent shaping of the expanding envelope 
by the radiation field and wind from the central star plays the lesser role 
of amplifying these features.  This conclusion is consistent with data from 
young Galactic PNe, except that for LMC PNe the determination of the size 
of the ionized nebula and the dynamical age are far more secure. 

We will present measurements and analysis of the nebular plasma diagnostics 
and chemical abundances in a future paper.  We will also present continuum 
magnitudes for the detected central stars, which will allow us to determine 
their evolutionary state.  These analyses form a key part of our intended 
research in this area, where we will be able to analyze the co-evolution of 
the PNe and their central stars without the debilitating uncertainties of 
the Galactic PN distance scale.  It will also give us a much more accurate 
picture of the chemical yields of PNe in the LMC.  We plan to extend 
this work to SMC PNe, where the lower metallicity environment should yield a 
difference in the morphological types, and perhaps also a difference in 
important aspects of the stellar and nebular evolution.  

\acknowledgements 

Support for this work was provided by NASA through grant number GO-08271.01-97A
from Space Telescope Science Institute, which is operated by the 
Association of Universities for Research in Astronomy, Incorporated, 
under NASA contract NAS5--26555.  We are grateful to an anonymous referee, 
whose comments helped us to improve this paper.

\clearpage

\centerline{\bf Figure Captions}

\figcaption[img1.ps]{
Images ({\it left}) and contour maps ({\it right}) of the LMC PNe 
J~41 ({\it upper}), SMP~4 ({\it middle}), and SMP~9 ({\it lower}). 
The images were obtained using the clear (50CCD) bandpass, and are 
3\arcsec\ on a side, with a log intensity stretch.  
The contour maps are from the G430M slit-less spectra, in the 
monochromatic light of [\ion{O}{3}] 5007\AA\ (except for J~41, which is 
in the light of H$\alpha$, taken from the G750M spectrum). 
The the contours are drawn at roughly 10\% intensity intervals. 
The orientation for each image is indicated on the figure, with north 
lying in the direction of the arrow, and east to the left. 
\label{Clear_1}}

\figcaption[img2.ps]{
Same as Fig.~\ref{Clear_1} for the nebulae: 
SMP~10 ({\it upper}), and SMP~16 ({\it lower}). 
These images are 4\arcsec\ on a side, but are on the same scale. 
\label{Clear_2}}

\figcaption[img3.ps]{
Same as Fig.~\ref{Clear_1} for the nebulae 
SMP~13 ({\it upper}), SMP~18 ({\it middle}), and SMP~19 ({\it lower}). 
\label{Clear_3}}

\figcaption[img4.ps]{
Same as Fig.~\ref{Clear_1} for the nebulae 
SMP~25 ({\it upper}), SMP~28 ({\it middle}), and SMP~30 
({\it lower}). 
\label{Clear_4}}

\figcaption[img5.ps]{
Same as Fig.~\ref{Clear_1} for the inner ({\it upper}) and outer ({\it lower}) 
portions of the nebula SMP~27.  The outer image is 9\arcsec\ by 12\arcsec\ 
and is presented at half the scale of the other images. 
\label{Clear_5}}

\figcaption[img6.ps]{
Same as Fig.~\ref{Clear_1} for the nebulae 
SMP~31 ({\it upper}), SMP~34 ({\it middle}), and SMP~46 ({\it lower}). 
\label{Clear_6}}

\figcaption[img7.ps]{
Same as Fig.~\ref{Clear_1} for the nebulae 
SMP~53 ({\it upper}), SMP~65 ({\it middle}), and SMP~71 ({\it lower}). 
\label{Clear_7}}

\figcaption[img8.ps]{
Same as Fig.~\ref{Clear_1} for the nebula SMP~59, except that the contour 
map is from the broadband (50CCD) image and the contours are evenly spaced 
in the square-root of the image intensity.  
This image is 5\arcsec\ square. 
\label{Clear_8}}

\figcaption[img9.ps]{
Same as Fig.~\ref{Clear_1} for the nebulae 
SMP~78 ({\it upper}), SMP~79 ({\it middle}), and SMP~80 ({\it lower}). 
\label{Clear_9}}

\figcaption[img10.ps]{
Same as Fig.~\ref{Clear_1} for the nebula SMP~93, except that the contour 
map is from the broadband (50CCD) image.  
This image is 6\arcsec\ by 8\arcsec\ and is at 75\% of the scale of the 
other images. 
\label{Clear_10}}

\figcaption[img11.ps]{
Same as Fig.~\ref{Clear_1} for the nebulae 
SMP~95 ({\it upper}), SMP~100 ({\it middle}), and SMP~102 ({\it lower}). 
\label{Clear_11}}

\clearpage

%
%

\begin{deluxetable}{lcllrc}
\tabletypesize{\scriptsize}
\tablewidth{0pt}
\tablecaption {Observing Log for LMC Planetary Nebulae \label{ObsLog}}
\tablehead {
\colhead {} & \colhead {} & \colhead {} & 
\colhead {} & \colhead {T$_{Exp}$} & \colhead {} \\
\colhead {Nebula} & \colhead {Date} & \colhead {Dataset} & 
\colhead {Disperser} & \colhead {(s)} & \colhead {N$_{Exp}$} 
}
\startdata  
J 41   & 1999-Sep-02 & O5BM50010 & MIRVIS &  300 & 2 \\*
       &             & O5BM50020 &  G750M & 1780 & 2 \\
SMP 4  & 1999-Aug-20 & O5BM22010 & MIRVIS & 120 & 2 \\*
       &             & O5BM22020 &  G750M & 730 & 2 \\*
       &             & O5BM22QHQ &  G750M &  90 & 1 \\*
       &             & O5BM22030 &  G430M & 280 & 2 \\
SMP 9  & 1999-Aug-12 & O5BM23010 & MIRVIS & 120 & 2 \\*
       &             & O5BM23020 &  G750M & 550 & 2 \\*
       &             & O5BM23LRQ &  G750M &  70 & 1 \\*
       &             & O5BM23030 &  G430M & 300 & 2 \\
SMP 10 & 1999-Sep-28 & O5BM24010 & MIRVIS & 120 & 2 \\*
       &             & O5BM24020 &  G750M & 360 & 2 \\*
       &             & O5BM24D2Q &  G750M &  45 & 1 \\*
       &             & O5BM24030 &  G430M & 360 & 2 \\
SMP 13 & 1999-Aug-15 & O5BM04010 & MIRVIS & 120 & 2 \\*
       &             & O5BM04BBQ & MIRVIS &  15 & 1 \\*
       &             & O5BM04020 &  G750M & 170 & 2 \\*
       &             & O5BM04BFQ &  G750M &  20 & 1 \\*
       &             & O5BM04030 &  G430M &  74 & 2 \\
SMP 16 & 1999-Aug-09 & O5BM27010 & MIRVIS &  300 & 2 \\*
       &             & O5BM27020 &  G750M & 1500 & 2 \\*
       &             & O5BM27030 &  G430M &  600 & 2 \\
SMP 18 & 1999-Aug-09 & O5BM29010 & MIRVIS & 120 & 2 \\*
       &             & O5BM29020 &  G750M & 580 & 2 \\*
       &             & O5BM29DZQ &  G750M &  72 & 1 \\*
       &             & O5BM29030 &  G430M & 340 & 2 \\
SMP 19 & 1999-Aug-20 & O5BM05010 & MIRVIS & 120 & 2 \\*
       &             & O5BM05QOQ & MIRVIS &  15 & 1 \\*
       &             & O5BM05020 &  G750M & 136 & 2 \\*
       &             & O5BM05QRQ &  G750M &  17 & 1 \\*
       &             & O5BM05030 &  G430M &  56 & 2 \\
SMP 25 & 1999-Aug-25 & O5BM06010 & MIRVIS & 120 & 2 \\*
       &             & O5BM06FWQ & MIRVIS &  15 & 1 \\*
       &             & O5BM06030 &  G750M & 300 & 2 \\*
       &             & O5BM06020 &  G750M &  60 & 2 \\*
       &             & O5BM06FZQ &  G750M &   8 & 1 \\*
       &             & O5BM06050 &  G430M & 170 & 2 \\*
       &             & O5BM06040 &  G430M &  34 & 1 \\
SMP 27 & 1999-Jul-19 & O5BM30010 & MIRVIS & 120 & 2 \\*
       &             & O5BM30020 &  G750M & 630 & 2 \\*
       &             & O5BM30DIQ &  G750M &  80 & 1 \\*
       &             & O5BM30030 &  G430M & 360 & 2 \\
SMP 28 & 1999-Sep-12 & O5BM31010 & MIRVIS & 120 & 2 \\*
       &             & O5BM31020 &  G750M & 560 & 2 \\*
       &             & O5BM31LCQ &  G750M &  70 & 1 \\*
       &             & O5BM31030 &  G430M & 280 & 2 \\
SMP 30 & 1999-Aug-17 & O5BM32010 & MIRVIS & 120 & 2 \\*
       &             & O5BM32020 &  G750M & 700 & 2 \\*
       &             & O5BM32HAQ &  G750M &  90 & 1 \\*
       &             & O5BM32030 &  G430M & 500 & 2 \\
SMP 31 & 1999-Aug-20 & O5BM33010 & MIRVIS &  120 & 2 \\*
       &             & O5BM33020 &  G750M &  210 & 2 \\*
       &             & O5BM33Q9Q &  G750M &   26 & 1 \\*
       &             & O5BM33030 &  G430M & 1120 & 2 \\
SMP 34 & 1999-Sep-27 & O5BM08010 & MIRVIS & 120 & 2 \\*
       &             & O5BM08020 &  G750M & 200 & 2 \\*
       &             & O5BM08CBQ &  G750M &  25 & 1 \\*
       &             & O5BM08030 &  G430M & 170 & 2 \\
SMP 46 & 1999-Sep-27 & O5BM36010 & MIRVIS &  120 & 2 \\*
       &             & O5BM36020 &  G750M & 1000 & 2 \\*
       &             & O5BM36030 &  G430M &  300 & 2 \\
SMP 53 & 1999-Sep-18 & O5BM11010 & MIRVIS & 120 & 2 \\*
       &             & O5BM11ZAQ & MIRVIS &  15 & 1 \\*
       &             & O5BM11020 &  G750M & 106 & 2 \\*
       &             & O5BM11ZDQ &  G750M &  13 & 1 \\*
       &             & O5BM11030 &  G430M &  38 & 2 \\
SMP 58 & 2000-Apr-29 & O5BM12010 & MIRVIS & 120 & 2 \\*
       &             & O5BM12CBQ & MIRVIS &  15 & 1 \\*
       &             & O5BM12030 &  G750M & 380 & 2 \\*
       &             & O5BM12020 &  G750M &  76 & 2 \\*
       &             & O5BM12CEQ &  G750M &  10 & 1 \\*
       &             & O5BM12040 &  G430M &  15 & 2 \\

SMP 59 & 1999-Aug-02 & O5BM39010 & MIRVIS & 120 & 2 \\*
       &             & O5BM39020 &  G750M & 880 & 2 \\*
       &             & O5BM39030 &  G430M & 480 & 2 \\
SMP 65 & 1999-Aug-31 & O5BM40010 & MIRVIS & 120 & 2 \\*
       &             & O5BM40020 &  G750M & 520 & 2 \\*
       &             & O5BM40EOQ &  G750M &  65 & 1 \\*
       &             & O5BM40030 &  G430M & 420 & 2 \\
SMP 71 & 1999-Aug-25 & O5BM13010 & MIRVIS & 120 & 2 \\*
       &             & O5BM13020 &  G750M & 180 & 2 \\*
       &             & O5BM13GCQ &  G750M &  22 & 1 \\*
       &             & O5BM13030 &  G430M &  70 & 2 \\
SMP 78 & 1999-Sep-28 & O5BM16010 & MIRVIS & 120 & 2 \\*
       &             & O5BM16FJQ & MIRVIS &  15 & 1 \\*
       &             & O5BM16030 &  G750M & 480 & 2 \\*
       &             & O5BM16020 &  G750M &  96 & 2 \\*
       &             & O5BM16FMQ &  G750M &  12 & 1 \\*
       &             & O5BM16040 &  G430M &  90 & 2 \\
SMP 79 & 1999-Sep-28 & O5BM17010 & MIRVIS & 120 & 2 \\*
       &             & O5BM17G6Q & MIRVIS &  15 & 1 \\*
       &             & O5BM17020 &  G750M & 108 & 2 \\*
       &             & O5BM17G9Q &  G750M &  13 & 1 \\*
       &             & O5BM17030 &  G430M & 120 & 2 \\
SMP 80 & 1999-Sep-29 & O5BM42010 & MIRVIS & 120 & 2 \\*
       &             & O5BM42020 &  G750M & 360 & 2 \\*
       &             & O5BM42KTQ &  G750M &  45 & 1 \\*
       &             & O5BM42030 &  G430M & 280 & 2 \\
SMP 81 & 1999-Aug-10 & O5BM18010 & MIRVIS & 120 & 2 \\*
       &             & O5BM18E6Q & MIRVIS &  15 & 1 \\*
       &             & O5BM18020 &  G750M & 103 & 2 \\*
       &             & O5BM18E9Q &  G750M &  13 & 1 \\*
       &             & O5BM18030 &  G430M &  33 & 2 \\
SMP 93 & 1999-Aug-11 & O5BM45010 & MIRVIS & 120 & 2 \\*
       &             & O5BM45020 &  G750M & 580 & 2 \\*
       &             & O5BM45LFQ &  G750M &  72 & 1 \\*
       &             & O5BM45030 &  G430M & 530 & 2 \\
SMP 94 & 1999-Jul-31 & O5BM20010 & MIRVIS & 120 & 2 \\*
       &             & O5BM20020 &  G750M & 250 & 2 \\*
       &             & O5BM20RAQ &  G750M &  31 & 1 \\*
       &             & O5BM20030 &  G430M &  90 & 2 \\*
       &             & O5BM20040 &  G430M & 480 & 2 \\
SMP 95 & 1999-Jul-31 & O5BM46010 & MIRVIS & 120 & 2 \\*
       &             & O5BM46020 &  G750M & 680 & 2 \\*
       &             & O5BM46QRQ &  G750M &  85 & 1 \\*
       &             & O5BM46030 &  G430M & 300 & 2 \\
SMP 100 & 1999-Aug-22 & O5BM21010 & MIRVIS & 120 & 2 \\*
        &             & O5BM21XTQ & MIRVIS &  15 & 1 \\*
        &             & O5BM21030 &  G750M & 240 & 2 \\*
        &             & O5BM21020 &  G750M &  42 & 1 \\*
        &             & O5BM21XWQ &  G750M &   5 & 2 \\*
        &             & O5BM21040 &  G430M & 150 & 2 \\
SMP 102 & 1999-Sep-29 & O5BM47010 & MIRVIS & 120 & 2 \\*
        &             & O5BM47020 &  G750M & 420 & 2 \\*
        &             & O5BM47G1Q &  G750M &  52 & 1 \\*
        &             & O5BM47030 &  G430M & 240 & 2 \\
\enddata  
\end{deluxetable}

%

\begin{deluxetable}{lrrclll}
\tabletypesize{\scriptsize}
\rotate  
\tablewidth{0pt}
\tablecaption {Positions, Dimensions and Morphologies of LMC Planetary Nebulae \label{Morph}}
\tablehead {
\colhead {} & \colhead {R.A.} & \colhead {Dec.} & \colhead {R$_{Phot}$} & 
\colhead {Dimensions} & \colhead {Morphological} & \colhead {} \\
\colhead {Nebula} & \colhead {(J2000)} & \colhead {(J2000)} & 
\colhead {(arcsec)} & \colhead {(arcsec)} & \colhead {Classification} & 
\colhead {Notes} 
}
\startdata  
J 41   &  5:26:09.51 &  $-69$:00:58.4 & 0.51 & inner: 0.69 x 0.62 & E(bc) & 
No [O III] image available; \\*
       &            &                 &      & outer: 1.17 x 1.10 &      & faint outer halo \\
SMP 4  &  4:43:21.50 &  $-71$:30:09.5 & 0.69 & 1.21        & E & Faint, round outer halo \\
SMP 9  &  4:50:24.71 &  $-68$:13:17.0 & 0.46 & 0.92 x 0.73 & E(bc) & Barrel shape \\
SMP 10 &  4:51:08.90 &  $-68$:49:05.8 & 0.88 & 1.58\tablenotemark{a} & P & Spiral shape \\
SMP 13 &  5:00:00.07 &  $-70$:27:41.8 & 0.47 & 0.81        & R(bc) &  \\
SMP 16 &  5:02:01.91 &  $-69$:48:54.4 & 1.00 & 1.5 x 1.2   & B & Butterfly shape \\
SMP 18 &  5:03:42.64 &  $-70$:06:47.8 & 0.51 & 0.69 x 0.64 & R(bc?) &  \\
SMP 19 &  5:03:41.30 &  $-70$:13:53.6 & 0.41 & 0.79 x 0.65 & E(bc) & Ring\\
SMP 25 &  5:06:24.00 &  $-69$:03:19.2 & 0.23 & 0.42 x 0.39 & R & \\
SMP 27 &  5:07:54.90 &  $-66$:57:46.1 & 0.46 & 0.76        & Q & Outer arc and blob\\
SMP 28 &  5:07:57.61 &  $-68$:51:47.0 & 0.41 & 0.58 x 0.35 & P & Outer contour non-E; arcs\\
SMP 30 &  5:09:10.61 &  $-66$:53:38.7 & 0.73 & 1.68 x 1.28 & B? & Very irregular\\
SMP 31 &  5:09:20.23 &  $-67$:47:25.2 & 0.15 & 0.26        & R? & \\
SMP 34 &  5:10:17.18 &  $-68$:48:23.0 & 0.32 & 0.57 x 0.50 & E & Slight asymmetry in [N II] \\
SMP 46 &  5:19:29.72 &  $-68$:51:09.1 & 0.31 & 0.59 x 0.49 & E(bc) & Ring?\\
SMP 53 &  5:21:32.93 &  $-67$:00:05.5 & 0.40 & 0.54 x 0.47 & E(bc?) & Barrel shape, could be B\\
SMP 58 &  5:24:20.81 &  $-70$:05:01.9 & 0.13 & 0.23        & R? &  \\
SMP 59 &  5:24:27.43 &  $-70$:22:24.7 & 1.50 & 3.70 x 2.66 & Q? &  \\
SMP 65 &  5:27:43.92 &  $-71$:25:56.6 & 0.37 &  0.59       & R &  \\
SMP 71 &  5:30:33.22 &  $-70$:44:38.4 & 0.31 & 0.58 x 0.47 & E &  \\
SMP 78 &  5:34:21.31 &  $-68$:58:24.7 & 0.28 & 0.54 x 0.42 & E(bc) & Barrel shape, could be B\\
SMP 79 &  5:34:08.76 &  $-74$:20:06.6 & 0.22 & 0.39 x 0.32 & E(bc) &  \\
SMP 80 &  5:34:38.87 &  $-70$:19:56.9 & 0.21 & 0.48        & R & Ring in [N II]\\
SMP 81 &  5:35:20.92 &  $-73$:55:30.1 & 0.15 & 0.26        & R? &  \\
SMP 93 &  5:49:38.80 &  $-69$:10:00.1 & 1.43 & 3.6 x 3.0   & B & Intersecting rings \\*
       &            &                 &      & 6.4 x 3.0   &   & Faint extensions \\
SMP 94 &  5:54:10.77 &  $-73$:02:47.5 & \nodata & $<0.1$   & \nodata & Unresolved; not a PN? \\
SMP 95 &  6:01:45.30 &  $-67$:56:08.0 & 0.46 & 1.15 x 0.95 & E(bc) & With ansae\\
SMP 100 & 6:22:55.73 &  $-72$:07:41.4 & 0.71 & 1.36 x 1.18 & E(bc) or Q? & \\
SMP 102 & 6:29:32.93 &  $-68$:03:32.9 & 0.71 & 1.06        & R(bc?) & \\
\enddata  
\tablenotetext{a}{Size excludes extended arms.}
\end{deluxetable}

%

\begin{deluxetable}{lcll}
\tablewidth{0pt}
\tablecaption {Dimensions and Morphologies of LMC PNe from WFPC2 \label{Dopita}}
\tablehead {
\colhead {} & 
\colhead {R$_{Phot}$} & \colhead {Dimensions} & \colhead {} \\
\colhead {Nebula} & 
\colhead {(arcsec)} & \colhead {(arcsec)} & \colhead {Morph. Class}  
}
\startdata  
SMP 1  & 0.18 & 0.33		& R \\
SMP 15 & 0.41 & 0.75 x 0.61 	& E(bc) \\
SMP 33 & 0.31 & 0.67 x 0.57	& E(bc) \\
SMP 38 & 0.25 & 0.57 x 0.40	& E \\
SMP 41 & 1.47 & 3.56 x 1.86	& B \\
SMP 42 & 0.46 & 0.83 x 0.67	& Q(ES) \\
SMP 50 & 0.35 & 0.68 x 0.61 	& E(bc?) \\
SMP 52 & 0.41 & 0.75	 	& R(bc?) \\
SMP 54 & 1.01 & 3.6 x 1.8	& B \\
SMP 55 & 0.18 & 0.36 		& R \\
SMP 56 & 0.32 & 0.55	 	& R \\
SMP 63 & 0.30 & 0.57 x 0.54 	& R \\
SMP 77 & 0.30 & 0.56 x 0.53 	& R \\
SMP 99 & 0.41 & 0.85 x 0.73	& E(bc) \\
\enddata  
\end{deluxetable}

%

\begin{deluxetable}{lcc}
\tablewidth{0pt}
\tablecaption {PN Morphological Types: LMC vs. Galaxy \label{Mtypes}}
\tablehead {
\colhead {Morphological} & \colhead {LMC} & \colhead {Galaxy\tablenotemark{a}} \\
\colhead {Class} & \colhead {(Percent)} & \colhead {(Percent)} 
}
\startdata  
Elliptical (E)		& 17 & 49 \\ 
Round (R)		& 29 & 23 \\ 
Bi-Polar (B)		& 10 & 14 \\ 
Bi-Polar Core (BC)	& 34 &  9 \\ 
Quadrupolar (Q)		&  7 &  3 \\ 
Point-Sym (P)		&  3 &  3 \\ 
Total, Asymmetric\tablenotemark{b} & 51 & 26 \\
\enddata  
\tablenotetext{a}{Derived from the sample defined in \citet{Manchado_etal00} 
and reclassified using the present scheme: see text.}

\tablenotetext{b}{Includes types B, BC, and Q.}
\end{deluxetable}

\end{document}